\input amstex
\magnification=\magstep1 
\baselineskip=13pt
\documentstyle{amsppt}
\vsize=8.7truein \CenteredTagsOnSplits \NoRunningHeads
\def\rk{\operatorname{rank}}
\def\sgn{\operatorname{sgn}}
\def\tr{\operatorname{tr}}
\def\zz{\bold{z}}
\def\per{\operatorname{per}}

\topmatter
\title  Stability and complexity of mixed discriminants \endtitle 
\author Alexander Barvinok  \endauthor
\address Department of Mathematics, University of Michigan, Ann Arbor,
MI 48109-1043, USA \endaddress
\email barvinok$\@$umich.edu  \endemail
\date April  2019 \enddate
\thanks  This research was partially supported by NSF Grant DMS 1361541.
\endthanks 
\abstract We show that the mixed discriminant of $n$ positive semidefinite $n \times n$ real symmetric matrices can be approximated within a relative error $\epsilon >0$ in quasi-polynomial $n^{O(\ln n -\ln \epsilon)}$ time, provided the distance of each matrix to the identity matrix in the operator norm does not exceed some absolute constant $\gamma_0 >0$. We deduce a similar result for the mixed discriminant of doubly stochastic $n$-tuples of matrices from the Marcus - Spielman - Srivastava bound on the roots of the mixed characteristic polynomial. Finally, we construct a quasi-polynomial algorithm for approximating the sum of $m$-th powers of principal minors of a matrix, provided the operator norm of the matrix is strictly less than 1. As is shown by Gurvits, for $m=2$ the problem is $\#P$-hard and covers the problem of computing the mixed discriminant of positive semidefinite matrices of rank 2.
\endabstract

\endtopmatter

\document

\head 1. Introduction and main results \endhead

Mixed discriminants were introduced by A.D. Alexandrov \cite{Al38} in his work on mixed volumes and what was later called ``the Alexandrov-Fenchel inequality".
Mixed discriminants generalize permanents and also found independent applications in problems of combinatorial counting, see, for example, Chapter 5 of \cite{BR97}, as well as in determinantal 
point processes \cite{C+17}, \cite{KT12}. Recently, they made a spectacular appearance in the ``mixed characteristic polynomial" introduced by Marcus, Spielman and Srivastava in their solution of the Kadison-Singer problem \cite{M+15}. Over the years, the problem of computing or approximating mixed discriminants efficiently attracted some attention \cite{GS02}, \cite{Gu05}, \cite{CP16}.

In this paper, we establish some stability properties of mixed discriminants (the absence of zeros in certain complex domains) and, as a corollary, construct efficient algorithms to approximate the mixed discriminant of some sets of matrices. For example, we show that the mixed discriminant of $n \times n$ positive semidefinite matrices
can be approximated within a relative error $\epsilon >0$ in quasi-polynomial $n^{O(\ln n -\ln \epsilon)}$ time, provided the distance of each matrix to the identity matrix in the operator norm does not exceed some absolute constant $\gamma_0 >0$. We also consider the case when all matrices have rank 2, shown to be $\#P$-hard by Gurvits \cite{Gu05}, and provide a quasi-polynomial approximation algorithm in a particular situation.

\subhead (1.1) Definitions and properties \endsubhead Let $A_1, \ldots, A_n$ be an $n$-tuple of $n \times n$ complex matrices. The {\it mixed discriminant} of $A_1, \ldots, A_n$ is defined by
$$D\left(A_1, \ldots, A_n \right)={\partial^n \over \partial t_1 \ldots \partial t_n} \det \left(t_1 A_1 + \ldots + t_n A_n \right).$$
The determinant in the right hand side is a homogeneous polynomial of degree $n$ in complex variables $t_1, \ldots, t_n$ and $D\left(A_1, \ldots, A_n\right)$ is the coefficient of the monomial $t_1 \cdots t_n$. It is not hard to see that $D\left(A_1, \ldots, A_n\right)$ is a polynomial of degree $n$ in the entries of $A_1, \ldots, A_n$: assuming that 
$A_k=\left(a_{ij}^k\right)$ for $k=1, \ldots, n$, we have 
$$D\left(A_1, \ldots, A_n\right) = \sum_{\sigma, \tau \in S_n} \sgn \sigma \prod_{i=1}^n a^{\tau(i)}_{i \sigma(i)} =
\sum_{\sigma, \tau \in S_n} \sgn (\sigma \tau) \prod_{i=1}^n a^i_{\tau(i) \sigma(i)}  , \tag1.1.1$$
where $S_n$ is the symmetric group of permutations of $\{1, \ldots, n\}$. 
It follows from (1.1.1) that $D\left(A_1, \ldots, A_n \right)$ is linear in each argument $A_i$, 
and symmetric under permutations of $A_1, \ldots, A_n$,
see, for example, Section 4.5 of \cite{Ba16}.

Mixed discriminants appear to be the most useful when the matrices $A_1, \ldots, A_n$ are positive semidefinite real symmetric (or complex Hermitian), in which case \newline $D\left(A_1, \ldots, A_n\right) \geq 0$.
For a real $n$-vector $x=\left(\xi_1, \ldots, \xi_n\right)$, let $x\otimes x$ denote the $n\times n$ matrix with the $(i, j)$-th entry equal $\xi_i \xi_j$. It is then not hard to see that 
$$D\left(x_1 \otimes x_1, \ldots, x_n \otimes x_n \right)=\left(\det \left[ x_1, \ldots, x_n \right]\right)^2, \tag1.1.2$$
where $\left[ x_1, \ldots, x_n\right]$ is the $n \times n$ matrix with columns $x_1, \ldots, x_n$, see, for example, Section 4.5 of \cite{Ba16}. Various applications of 
mixed discriminants are based on (1.1.2). Suppose that $S_1, \ldots, S_n \subset {\Bbb R}^n$ are finite sets of vectors. Let us define 
$$A_k=\sum_{x \in S_k} x \otimes x \quad \text{for} \quad k=1, \ldots, n.$$
From the linearity of $D\left(A_1, \ldots, A_n\right)$ in each argument, we obtain 
$$D\left(A_1, \ldots, A_n\right) = \sum_{x_1 \in S_1, \ldots, x_n \in S_n}  \left( \det\left[ x_1, \ldots, x_n\right]\right)^2. \tag1.1.3$$
One combinatorial application of (1.1.3) is as follows: given a connected graph $G$ with $n$ vertices, color the edges of $G$ in $n-1$ colors. Then the number of spanning trees containing exactly one edge of each color is naturally expressed as a mixed discriminant. More generally, this extends to counting ``rainbow bases" in regular matroids
with colored elements, cf. 
Chapter 5 of \cite{BR97}. Another application of (1.1.3) is in determinantal point processes \cite{C+17}.

The mixed discriminant of (positive semidefinite) matrices generalizes the permanent of a (non-negative) matrix. Namely, given $n \times n$ diagonal matrices $A_1, \ldots, A_n$, we consider an $n \times n$ matrix $B$ whose $k$-th row is the diagonal of $A_k$. It is easy to see that 
$$D\left(A_1, \ldots, A_n\right) = \per B,$$
where the {\it permanent} of $B$ is defined by 
$$\per B = \sum_{\sigma \in S_n} \prod_{i=1}^n b_{i \sigma(i)}.$$ We note that if $A_1, \ldots, A_n$ are positive semidefinite then $B$ is a non-negative matrix and that the permanent of any non-negative square matrix can be interpreted as the mixed discriminant of positive semidefinite matrices.

In their solution of the Kadison-Singer problem, Marcus, Spielman and Srivastava defined the {\it mixed characteristic polynomial} of $n \times n$ matrices 
$A_1, \ldots, A_m$ by 
$$p(x)=p_{A_1, \ldots, A_m}(x)=  \prod_{k=1}^m \left(1 -{\partial \over \partial z_k}\right) \det\left(xI + \sum_{k=1}^m z_k A_k \right)\Big|_{z_1= \ldots = z_m=0} \tag1.1.4$$
\cite{M+15}, see also \cite{MS17}.
The coefficients of $p(x)$ can be easily expressed as mixed discriminants, see Section 5.1. Finding efficiently the partition in the Weaver's reformulation of the Kadison-Singer conjecture (the existence of such a partition is proven in \cite{M+15}) reduces to bounding the roots of the mixed characteristic polynomial, which in turn makes computing the coefficients of the polynomial (which are expressed as mixed discriminants) of interest. It follows from our results that for any non-negative integer $c$ and $k$, fixed in advance, one can approximate the coefficient of $x^k$ in quasi-polynomial time, provided $0 \leq m - n \leq c$ and the distance from each matrix $A_i$ to $I$ in the operator norm does not exceed an absolute constant $\gamma_0 >0$.

Let $B=\left(b_{ij}\right)$ be an $n \times n$ complex matrix. For a set $S \subset \{1, \ldots, n\}$, let $B_S$ be the $|S| \times |S|$ submatrix of $B$ consisting of the entries $b_{ij}$ with 
$i, j \in S$. Gurvits \cite{Gu05} noticed that computing the mixed discriminant of positive semidefinite matrices $A_1, \ldots, A_n$ of rank 2 reduces to computing the sum
$$\sum_{S \subset \{1, \ldots, n\}} \left(\det B_S\right)^2,$$
where for $S=\emptyset$ the corresponding term is equal to 1. Indeed, applying a linear transformation if necessary, we assume that for $k=1, \ldots, n$ we have  $A_k =e_k \otimes e_k + x_k \otimes x_k$, where $e_1, \ldots, e_n$ is the standard basis of ${\Bbb R}^n$ and $x_1, \ldots, x_n$ are some vectors in ${\Bbb R}^n$.
From the linearity of the mixed discriminant in each argument and (1.1.2), it follows that 
$$D\left(e_1 \otimes e_1 + x_1 \otimes x_1, \ldots, e_n \otimes e_n + x_n \otimes x_n \right) = \sum_{S \subset \{1, \ldots, n\}} \left(\det B_S \right)^2,$$
where $B=\left(b_{ij}\right)$ is the Gram matrix of vectors $x_1, \ldots, x_n$, that is, $b_{ij}=\langle x_i, x_j \rangle$ and $\langle \cdot, \cdot \rangle$ is the standard scalar product in 
${\Bbb R}^n$, see \cite{Gu05} for details. Computing a more general expression 
$$\sum_{S \subset \{1, \ldots, n \}} \left(\det B_S \right)^m \tag1.1.5$$ 
for an integer $m \geq 2$ is of interest in discrete determinantal point processes \cite{KT12}.

As a ramification of our approach, we present a quasi-polynomial algorithm of $n^{O(\ln n - \ln \epsilon)}$ complexity approximating (1.1.5) within relative error $0 < \epsilon <1$
provided $\|B\| < \rho$ for any $\rho < 1$, fixed in advance, where $\| \cdot \|$ is the operator norm.

\subhead (1.2) Computational complexity \endsubhead Since mixed discriminants generalize permanents, they are at least as hard to compute exactly or to approximate as permanents. Moreover, it appears that mixed discriminants are substantially harder to deal with than permanents. It is shown in \cite{Gu05} that it is a $\#P$-hard problem to compute 
$D(A_1, \ldots, A_n)$ even when $\rk A_k=2$ for $k=1, \ldots, n$. In particular, computing (1.1.5) for a positive definite matrix $B$ and $m=2$ is a $\#P$-hard problem.
In contrast, the permanent of a matrix with at most 2 non-zero entries in each row is trivial to compute. 
A Monte Carlo Markov Chain algorithm of Jerrum, Sinclair and Vigoda \cite{J+04} approximates the permanent of a non-negative matrix in randomized polynomial time. 
Nothing similar is known or even conjectured to work for the mixed discriminant of positive semidefinite matrices. A randomized polynomial time algorithm from \cite{Ba99} approximates the mixed discriminant of $n \times n$ positive semidefinite matrices within a multiplicative factor of $c^n$ for $c=2 e^{\gamma-1} \approx 1.31$, where 
$\gamma \approx 0.577$ is the Euler constant. A deterministic polynomial time algorithm of \cite{GS02} approximates the mixed discriminants of positive semidefinite matrices within a multiplicative factor of $e^n \approx (2.718)^n$.  As is shown in Section 4.6 of \cite{Ba16}, for any $\gamma >1$, fixed in advance, the scaling algorithm of \cite{GS02} approximates the mixed discriminant $D\left(A_1, \ldots, A_n\right)$ within a multiplicative factor of 
$n^{\gamma^2}$ provided the largest eigenvalue of each matrix $A_k$ is within a factor of $\gamma$ of its smallest eigenvalue.

A combinatorial algorithm of \cite{CP16} computes the mixed discriminant exactly in polynomial time for some class of matrices (of bounded tree width).

Our first result establishes the absence of complex zeros of $D\left(A_1, \ldots, A_n\right)$ if all $A_i$ lie sufficiently close to the identity matrix. In what follows, $\| \cdot\|$ denotes 
the operator norm a matrix, which in the case of a real symmetric matrix is the largest absolute value of an eigenvalue.
\proclaim{(1.3) Theorem} There is an absolute constant $\gamma_0 >0$ (one can choose $\gamma_0=0.045$) such that if $Q_1, \ldots, Q_n$ are $n \times n$ real symmetric 
matrices satisfying 
$$\|Q_k\| \ \leq \ \gamma_0 \quad \text{for} \quad k=1, \ldots, n$$
then for $z_1, \ldots, z_n \in {\Bbb C}$, we have 
$$D\left(I +z_1 Q_1, \ldots, I + z_n Q_n \right) \ne 0 \quad \text{provided} \quad |z_1|, \ldots, |z_n| \leq 1.$$
\endproclaim
We note that under the conditions of the theorem, the mixed discriminant is not confined to any particular sector of the complex plane (in other words, the reasons for the mixed discriminant to be non-zero are not quite straightforward). For example, if $Q_1=\ldots =Q_n =\gamma_0 I$, then $D\left(I+ z Q_1, \ldots, I + zQ_n\right)=n! (1+\gamma_0 z)^n$ rotates 
$\Omega(n)$ times around the origin as $z$ ranges over the unit circle.

Applying the interpolation technique, see \cite{Ba16}, \cite{PR17}, we deduce that the mixed discriminant $D\left(A_1, \ldots, A_n\right)$ can be efficiently approximated if the 
matrices \newline $A_1, \ldots, A_n$ are close to $I$ in the operator norm. Let $Q_1, \ldots, Q_n$ be matrices satisfying the conditions of Theorem 1.3. Since 
$D\left(I+z_1 Q_1, \ldots, I + z_n Q_n \right) \ne 0$ in the simply connected domain (polydisc) $|z_1|, \ldots, |z_n| \leq 1$, we can choose a branch of 
$\ln D\left(I + z_1 Q_1, \ldots, I + z_n Q_n\right)$ in that domain. It turns out that the logarithm of the mixed discriminant can be efficiently approximated by a low (logarithmic) degree polynomial.

\proclaim{(1.4) Theorem} For any $0 < \rho < 1$ there is a constant $c(\rho) >0$ and for any $0 < \epsilon <1$, for any positive integer $n$, there is a polynomial 
$$p=p_{\rho, n, \epsilon}\left(Q_1, \ldots, Q_n; z_1, \ldots, z_n \right)$$ in the entries 
of $n \times n$ real symmetric matrices $Q_1, \ldots, Q_n$ and complex $z_1, \ldots, z_n$ such that $\deg p \ \leq \ c(\rho) \left( \ln n - \ln \epsilon \right)$ and 
$$\left| \ln D\left(I + z_1 Q_1, \ldots, I + z_n Q_n \right) - p(Q_1, \ldots, Q_n; z_1, \ldots, z_n)\right| \leq \epsilon$$
provided 
$$\|Q_k\| \ \leq \ \gamma_0 \quad \text{for} \quad k=1, \ldots, n$$
where $\gamma_0$ is the constant in Theorem 1.4 and 
$$|z_1|, \ldots, |z_n| \ \leq \ \rho.$$
\endproclaim
We show that the polynomial $p$ can be computed in quasi-polynomial \newline $n^{O_{\rho}(\ln n - \ln \epsilon)}$ time, where the implicit constant in the ``$O$" notation depends on $\rho$ alone. In other words, Theorem 1.4 implies that the mixed discriminant of positive definite matrices $A_1, \ldots, A_n$ can be approximated within a relative error $\epsilon >0$ in quasi-polynomial  $n^{O(\ln n - \ln \epsilon)}$ time provided for each matrix $A_k$, the ratio of any two eigenvalues is bounded by a 
constant $1< \gamma < (1+\gamma_0)/(1-\gamma_0)$, fixed in advance.
We note that the mixed discriminant of such $n$-tuples can vary within an exponentially large multiplicative factor $\gamma^n$.

Theorem 1.4 shows that the mixed discriminant can be efficiently approximated in some open domain in the space of $n$-tuples of $n \times n$ symmetric matrices. 
A standard argument shows that unless $\#P$-hard problems can be solved in quasi-polynomial time, the mixed discriminant cannot be computed {\it exactly} in any open domain in quasi-polynomial time: if such a domain existed, we could compute the mixed discriminant exactly at any $n$-tuple as follows: we choose a line through the desired $n$-tuple and an $n$-tuple in the domain; since the restriction of the mixed discriminant onto a line is a polynomial of degree $n$, we could compute it by interpolation from the values at points in the domain.

We deduce from the Marcus - Spielman - Srivastava bound on the roots of the mixed characteristic polynomial \cite{MS17} the following stability result for mixed discriminants. 
\proclaim{(1.5) Theorem} Let $\alpha_0\approx 0.278$ be the positive real solution of the equation $\alpha e^{1+\alpha}=1$. Suppose that $Q_1, \ldots, Q_n$ are $n \times n$ positive semidefinite matrices such that $Q_1 + \ldots + Q_n=I$ and $\tr Q_k=1$ for $k=1, \ldots, n$. Then 
$$D\left(I +z Q_1, \ldots, I + zQ_n\right) \ne 0 \quad \text{for all $z \in {\Bbb C}$ such that} \quad |z| < {\alpha_0 n \over 4}.$$
\endproclaim 

As before, the interpolation argument produces the following algorithmic corollary.
\proclaim{(1.6) Theorem} For any $0 < \rho < \alpha_0/4 \approx 0.072$, where $\alpha_0$ is the constant of Theorem 1.5, there is a constant $c(\rho)>0$, and for any $0 < \epsilon < 1$, and any positive integer $n$ 
there is a polynomial 
$$p=p_{\rho, n, \epsilon}\left(Q_1, \ldots, Q_n; z \right)$$ in the entries of $n \times n$ real symmetric matrices $Q_1, \ldots, Q_n$ and complex $z$ such that 
$\deg p \ \leq \ c(\rho) \left( \ln n - \ln \epsilon\right)$ and
$$\left| \ln D\left(I  +z   Q_1, \ldots, I +z Q_n \right) - p\left(Q_1, \ldots, Q_n; z \right) \right| \ \leq \ \epsilon$$
provided $Q_1, \ldots, Q_n$ are $n \times n$ positive semidefinite matrices such that 
$$Q_1+ \ldots + Q_n=I, \quad \tr Q_k=1 \quad \text{for} \quad k=1, \ldots, n \tag1.6.1$$ and 
$|z| \leq \rho n$.
\endproclaim
Again, the polynomial $p$ is constructed in quasi-polynomial $n^{O(\ln n-\ln \epsilon)}$ time.

Some remarks are in order. An $n$-tuple of $n \times n$ positive semidefinite matrices $Q_1, \ldots, Q_n$ satisfying (1.6.1) is called {\it doubly stochastic}.
Gurvits and Samorodnitsky \cite{GS02} proved that an $n$-tuple $A_1, \ldots, A_n$ of $n \times n$ positive definite matrices can be {\it scaled} (efficiently, in polynomial times) to a doubly stochastic $n$-tuple, that is, one can find an $n \times n$ matrix $T$, a doubly stochastic $n$-tuple $Q_1, \ldots, Q_n$, and positive real numbers 
$\xi_1, \ldots, \xi_n$ such that $A_k=\xi_k T^{\ast} Q_k T$ for $k=1, \ldots, n$, see also Section 4.5 of \cite{Ba16} for an exposition. Then we have
$$D\left(A_1, \ldots, A_n\right) = \xi_1 \cdots \xi_n (\det T)^2 D\left(Q_1, \ldots, Q_n\right)$$
and hence computing the mixed discriminant for any $n$-tuple of positive semidefinite matrices reduces to that for a doubly stochastic $n$-tuple.
The $n$-tuple $C=\left(n^{-1} I, \ldots, n^{-1}I \right)$ naturally plays the role of the ``center" of the set of all doubly stochastic $n$-tuples. Let us contract the convex body of 
all doubly stochastic $n$-tuples $X$ towards its center $C$ with a constant coefficient $\gamma < \alpha_0/4 \approx 0.07$, 
$X \longmapsto (1-\gamma) C + \gamma X$. Theorem 1.5 implies that the mixed discriminants of all contracted $n$-tuples are efficiently (in quasi-polynomial time) approximable. In other words, there is ``core" of the convex body of doubly stochastic $n$-tuples, where the mixed discriminant is efficiently approximable, and that core is just a scaled copy (with a constant, small but positive, scaling coefficient) of the whole body.

Finally, we address the problem of computing (1.1.5). First, we prove the following stability result.
\proclaim{(1.7) Theorem} For an $n \times n$ complex matrix $B=\left(b_{ij}\right)$ and a set $S \subset \{1, \ldots, n\}$, let 
$B_S$ be the submatrix of $B$ consisting of the entries $b_{ij}$ with $i, j \in S$. For an integer $m \geq 1$, we define a polynomial
$$\phi_{B,m}(z)=\sum_{S \subset \{1, \ldots, n\}} \left( \det B_S \right)^m z^{|S|},$$
with constant term $1$ corresponding to $S=\emptyset$. If $\|B\| < 1$ then 
$$\phi_{B,m}(z) \ne 0 \quad \text{for any} \quad z \in {\Bbb C} \quad \text{such that} \quad |z| \leq 1.$$
\endproclaim

Consequently, by interpolation we obtain the following result.
\proclaim{(1.8) Theorem} For any $0 < \rho < 1$ and any integer $m \geq 1$ there is a constant $c(\rho, m) >0$ and for any $0 < \epsilon < 1$ and integer $n$ there is a polynomial 
$$p=p_{\rho, m, \epsilon, n}(B)$$
in the entries of an $n \times n$ complex matrix $B$ such that $\deg p \leq c(\rho, m)\left(\ln n - \ln \epsilon\right)$ and 
$$\left| \ln \left(\sum_{S \subset \{1, \ldots, n\}} \left( \det B_S \right)^m \right) - p(B) \right| \ \leq \ \epsilon,$$
provided $B$ is an $n \times n$ matrix such that $\|B\| < \rho$.
\endproclaim
The polynomial $p$ is constructed in quasi-polynomial $n^{O_{\rho, m}(\ln n -\ln \epsilon)}$ time.

We prove Theorem 1.3 in Sections 2 and 3. We prove Theorem 1.4 in Section 4. In Section 5, we prove Theorems 1.5 and 1.6 and in Section 6, we prove Theorems 1.7 and 1.8.

\head 2. Preliminaries \endhead

\subhead (2.1) From matrices to quadratic forms \endsubhead  Let $\langle \cdot, \cdot \rangle$ be the standard scalar product in ${\Bbb R}^n$. With an $n \times n$ real symmetric matrix $Q$ we associate a quadratic form $q: {\Bbb R}^n \longrightarrow {\Bbb R}$,
$$q(x)=\langle x, Q x \rangle \quad \text{for} \quad x \in {\Bbb R}^n.$$
Given quadratic forms $q_1, \ldots, q_n: {\Bbb R}^n \longrightarrow {\Bbb R}$, we define their mixed discriminant by
$$D\left( q_1, \ldots, q_n \right) = D\left(Q_1, \ldots, Q_n \right), $$
where $Q_k$ is the matrix of $q_k$. This definition does not depend on the choice of an orthonormal basis in ${\Bbb R}^n$ (as long as the scalar product remains fixed):
if we change the basis, the matrices change as $Q_k:= U^{\ast} Q_k U$ for some orthogonal matrix $U$ and all $k=1, \ldots, n$, and hence the mixed discriminant does not change.

The advantage of working with quadratic forms is that it allows us to define the mixed discriminant of the restriction of the forms onto a subspace. Namely, if 
$q_1, \ldots, q_m: {\Bbb R}^n \longrightarrow {\Bbb R}$ are quadratic forms and $L \subset {\Bbb R}^n$ is a subspace with $\dim L=m$, we make $L$ into a Euclidean space with the scalar product inherited from ${\Bbb R}^n$ and define the mixed discriminant $D\left(q_1|L, \ldots, q_m|L \right)$ for the restrictions $q_k: L \longrightarrow {\Bbb R}$.

We will use the following simple lemma.
\proclaim{(2.2) Lemma} Let $q_1, \ldots, q_n: {\Bbb R}^n \longrightarrow {\Bbb R}$ be quadratic forms and suppose that 
$$q_n(x)=\sum_{k=1}^n \lambda_k \langle u_k, x \rangle^2,$$
where $\lambda_1, \ldots, \lambda_n$ are real numbers and $u_1, \ldots, u_n \in {\Bbb R}^n$ are unit vectors.
Then 
$$D\left(q_1, \ldots, q_n\right)=\sum_{k=1}^n \lambda_k D\left(q_1|u_k^{\bot}, \ldots, q_{n-1}|u_k^{\bot} \right),$$
where $u_k^{\bot}$ is the orthogonal complement to $u_k$.
\endproclaim
\demo{Proof}  This is Lemma 4.6.3 from \cite{Ba16}. We give its proof here for completeness. 
By the linearity of the mixed discriminant in each argument, it suffices to check the formula when $q_n(x) = \langle u, x \rangle^2$, where $u$ is a unit vector. 
Let $Q_1, \ldots, Q_n$ be the matrices of $q_1, \ldots, q_n$ in an orthonormal basis, where $u$ is the $n$-th basis vector and hence $Q_n$ is the matrix where the $(n, n)$-th entry is $1$ and all other entries are $0$.

 It follows from (1.1.1) that 
 $$D\left(Q_1, \ldots, Q_n\right) =D\left(Q_1', \ldots, Q_{n-1}'\right),$$
 where $Q_k'$ is the upper left $(n-1) \times (n-1)$ submatrix of $Q_k$. We observe that $Q_k'$ is the matrix of the restriction $q_k|u^{\bot}$.
{\hfill \hfill \hfill} \qed
\enddemo

\subhead (2.3) Comparing two restrictions \endsubhead
Let $q_1, \ldots, q_{n-1}: {\Bbb R}^n \longrightarrow {\Bbb R}$ be quadratic forms and let $u, v \in {\Bbb R}^n$ be unit vectors (we assume that $u \ne \pm v)$. We would like to compare 
$D\left(q_1|u^{\bot}, \ldots, q_{n-1}|u^{\bot}\right)$ and $D\left(q_1|v^{\bot}, \ldots, q_{n-1}| v^{\bot}\right)$. Let $L = u^{\bot} \cap v^{\bot}$, so $L \subset {\Bbb R}^n$ is a subspace of codimension 2. Let us identify $u^{\bot}$ and $v^{\bot}$ with ${\Bbb R}^{n-1}$ as Euclidean spaces (we want to preserve the scalar product but do not worry about bases) in such a way that $L$ gets identified with ${\Bbb R}^{n-2} \subset {\Bbb R}^{n-1}$. 
Hence the quadratic forms $q_k|u^{\bot}$ get identified with some quadratic forms $q_k^u: {\Bbb R}^{n-1} \longrightarrow {\Bbb R}$ and the quadratic forms 
$q_k|v^{\bot}$ get identified with some quadratic forms $q_k^v: {\Bbb R}^{n-1} \longrightarrow {\Bbb R}$ for $k=1, \ldots, n-1$.

We have 
$$\split &D\left(q_1|u^{\bot}, \ldots, q_{n-1}|u^{\bot}\right) = D\left(q_1^u, \ldots, q_{n-1}^u \right) \quad \text{and} \\
&D\left(q_1|v^{\bot}, \ldots, q_{n-1}|v^{\bot}\right)=D\left(q_1^v, \ldots, q_{n-1}^v \right). \endsplit$$
Besides
$$q_k^u | {\Bbb R}^{n-2} = q_k^v|{\Bbb R}^{n-2}  \quad \text{for} \quad k=1, \ldots, n-1. \tag2.3.1$$
Let us denote 
$$r_k(x)=q_k^u(x)-q_k^v(x) \quad \text{for} \quad k=1, \ldots, n-1.$$
Hence $r_k: {\Bbb R}^{n-1} \longrightarrow {\Bbb R}$ are quadratic forms and by (2.3.1) we have $r_k(x)=0$ for all $x \in {\Bbb R}^{n-2}$. It follows then that 
$$\aligned &r_k(x) = \xi_{n-1} \ell_k(x) \quad \text{where} \quad x=\left(\xi_1, \ldots, \xi_{n-1}\right) \\ &\text{and} \quad \ell_k: {\Bbb R}^{n-1} \longrightarrow {\Bbb R} \quad  
\text{are linear forms} \quad  \text{for} \quad  k=1, \ldots, n-1. \endaligned \tag2.3.2$$

\proclaim{(2.4) Lemma} Suppose that $n \geq 3$. Let $w, \ell_1, \ell_2, \ell_3: {\Bbb R}^n \longrightarrow {\Bbb R}$ be linear forms. For $k=1, 2, 3$, let $r_k(x)=w(x) \ell_k(x)$ be quadratic forms and let $q_4, \ldots, q_n: {\Bbb R}^n \longrightarrow {\Bbb R}$ be some other quadratic forms. Then 
$$D\left(r_1, r_2, r_3, q_4, \ldots, q_n \right)=0.$$
\endproclaim
\demo{Proof} Since the restriction of a linear form onto a subspace is a linear form on the subspace, repeatedly applying Lemma 2.2, we reduce the general case to the case of $n=3$, in which case the mixed discriminant in question is just $D\left(r_1, r_2, r_3\right)$. On the other hand, for all real $t_1, t_2, t_3$ we have 
$$\rk \left( t_1 r_1 + t_2 r_2 + t_3 r_3\right)=\rk \left(w \cdot \left(t_1 \ell_1 + t_2 \ell_2 + t_3 \ell_3 \right)\right) \leq 2,$$
and hence 
$$\det   \left( t_1 r_1 + t_2 r_2 + t_3 r_3\right)=0.$$
It follows by Definition 1.1 that $D\left(r_1, r_2, r_3\right)=0$.
{\hfill \hfill \hfill} \qed
\enddemo

\proclaim{(2.5) Corollary} Suppose that $n \geq 2$ and let $q_1, \ldots, q_{n-1}: {\Bbb R}^n \longrightarrow {\Bbb R}$ be quadratic forms. Let $u, v \in {\Bbb R}^d$ be 
unit vectors such that $u \ne -v$ and for $k=1, \ldots, n-1$, let us define quadratic forms 
$q_k^u: {\Bbb R}^{n-1} \longrightarrow {\Bbb R}$ and $q_k^v: {\Bbb R}^{n-1} \longrightarrow {\Bbb R}$ as in Section 2.3.
Let $r_k=q_k^u- q_k^v$ for $k=1, \ldots, n-1$. Then 
$$\split D\left(q_1^u, \ldots, q_{n-1}^u \right)=&D\left(q_1^v, \ldots, q_{n-1}^v\right) + \sum_{i=1}^{n-1} D\left(r_i, q_k^v:\ k \ne i \right) \\&\quad + \sum_{1 \leq i < j \leq n-1}
D\left(r_i, r_j, q^v_k: \ k \ne i, j \right).\endsplit$$
Moreover,
$$\rk r_k \ \leq 2 \quad \text{for} \quad k=1, \ldots, n-1.$$
If $n=2$, the second sum in the right hand side is empty.
\endproclaim
\demo{Proof} Since $q_k^u = q_k^v +r_k$ for $k=1, \ldots, n-1$, the proof follows by the linearity of the mixed discriminant in each argument and by Lemma 2.4.
{\hfill \hfill \hfill} \qed
\enddemo

Finally, we will need a simple estimate.
\proclaim{(2.6) Lemma} Let $z$ and $w$ be complex numbers such that $|1-z| \leq \delta$ and $|1-w| \leq \tau$ for some $0 < \delta, \tau < 1$.
Then 
$$\Re \thinspace zw \ \geq \ 1- \delta - \tau - \delta \tau.$$
\endproclaim
\demo{Proof} We write $z=1+a e^{i \phi}$ and $w=1+b e^{i \psi}$ for some real $a, b, \phi$ and $\psi$ such that $0 \leq a \leq \delta$ and $0 \leq b \leq \tau$.
Then
$$\Re\thinspace zw =1 + a \cos \phi + b \cos \psi + ab \cos (\phi+\psi) \ \geq \ 1- a - b - ab \ \geq \ 1-\delta-\tau -\delta \tau.$$
{\hfill \hfill \hfill} \qed
\enddemo

\head 3. Proof of Theorem 1.3 \endhead

We prove Theorem 1.3 by induction on $n$. Following Section 2, we associate with (now complex) matrices $I+z_kQ_k$ (now complex-valued) quadratic 
forms $p_k(x)=\|x\|^2 +z_k q_k(x)$ where $q_k: {\Bbb R}^n \longrightarrow {\Bbb R}$ is the quadratic form with matrix $Q_k$. If $L \subset {\Bbb R}^n$ is a subspace then the restriction of $p_k$ onto $L$ is just $\|x\|^2 + z_k (q_k|L)$, where $q_k|L$ is the restriction of $q_k$ onto $L$. The induction is based on the following two lemmas.
\proclaim{(3.1) Lemma} Let us fix $0 < \delta, \tau < 1$ such that $\delta + \tau + \delta \tau < 1$. Let $q_1, \ldots, q_n: {\Bbb R}^n \longrightarrow {\Bbb R}$, $n \geq 2$, 
be quadratic forms and let $z_1, \ldots, z_n$ be complex numbers such that $|z_1|, \ldots, |z_n| \leq 1$. Let us define 
$$p_k(x) = \|x\|^2+z_k q_k(x), \quad p_k: {\Bbb R}^n \longrightarrow {\Bbb C} \quad \text{for} \quad k=1, \ldots, n$$ and 
suppose that the following conditions hold:
\roster
\item For any two unit vectors $u, v \in {\Bbb R}^n$ one can write
$$D\left(p_1|u^{\bot}, \ldots, p_{n-1}|u^{\bot}\right) = \left(1 + \alpha(u, v) \right) D\left(p_1|v^{\bot}, \ldots, p_{n-1}|v^{\bot} \right)$$
for some $\alpha(u, v) \in {\Bbb C}$ such that $|\alpha(u, v)| \leq \tau$;
\item We have 
$$|q_n(x)| \ \leq \delta \|x\|^2 \quad \text{for all} \quad x \in {\Bbb R}^n.$$
\endroster
Then for any unit vector $v \in {\Bbb R}^n$, we have 
$$\left|D\left(p_1, \ldots, p_n \right) \right| \ \geq \ n(1- \delta - \tau - \tau \delta) \left| D\left(p_1 |v^{\bot}, \ldots, p_{n-1}|v^{\bot} \right)\right|.$$
\endproclaim
\demo{Proof} We have 
$$q_n(x) =\sum_{k=1}^n \lambda_k \langle u_k, x \rangle^2,$$
where $u_1, \ldots, u_n$ are the orthonormal eigenvectors of $q_n$ and $\lambda_1, \ldots, \lambda_n$ are the corresponding eigenvalues. 
In particular, from condition (2) of the lemma, we have 
$$|\lambda_k| \ \leq \ \delta \quad \text{for} \quad k=1, \ldots, n.$$
Since 
$$\sum_{k=1}^n \langle u_k, x \rangle^2 = \|x\|^2,$$
from Lemma 2.2, we obtain by the linearity of the mixed discriminant
$$D\left(p_1, \ldots, p_n\right)=\sum_{k=1}^n \left(1+z_n \lambda_k \right) D\left(p_1|u_k^{\bot}, \ldots, p_{n-1}|u_k^{\bot}\right). \tag3.1.1$$
Let us choose a unit vector $v \in {\Bbb R}^n$. Then 
$$D\left(p_1|u_k^{\bot}, \ldots, p_{n-1}|u_k^{\bot}\right) = \left(1 + \alpha(u_k, v)\right) D\left(p_1|v^{\bot}, \ldots, p_{n-1}|v^{\bot}\right) \tag3.1.2$$
for some $\alpha(u_k, v) \in {\Bbb C}$ such that $|\alpha(u_k, v)| \leq \tau$.

Combining (3.1.1) and (3.1.2), we get 
$$D\left(p_1, \ldots, p_n\right) = D\left(p_1|v^{\bot}, \ldots, p_{n-1}|v^{\bot}\right) \sum_{k=1}^n \left(1 + z_n \lambda_k \right) \left(1 + \alpha(u_k, v)\right). \tag3.1.3$$
From Lemma 2.6, 
$$\Re \left( \sum_{k=1}^n \left(1 + z_n \lambda_k \right) \left(1+ \alpha(u_k, v) \right) \right) \ \geq \ n (1- \delta - \tau - \delta \tau) \ > \ 0$$ 
and hence 
$$\left| \sum_{k=1}^n \left(1 + z_n \lambda_k \right) \left(1 + \alpha(u_k, v)\right) \right| \ \geq \ n(1- \delta - \tau - \delta \tau).$$
The proof then follows from (3.1.3).
{\hfill \hfill \hfill} \qed
\enddemo

\proclaim{(3.2) Lemma} Let us fix $0 < \delta, \mu < 1$ such that 
$$4 \delta \mu^{-1} + 8 \delta^2 \mu^{-2} \ < \ 1.$$ Let $q_1, \ldots, q_{n-1}: {\Bbb R}^n \longrightarrow {\Bbb R}$, $n \geq 2$, be quadratic forms and let $z_1, \ldots, z_{n-1}$ be complex numbers such that $|z_1|, \ldots, |z_{n-1}| \leq 1$.  Let us define 
$$p_k(x) = \|x\|^2+z_k q_k(x), \quad p_k: {\Bbb R}^n \longrightarrow {\Bbb C} \quad \text{for} \quad k=1, \ldots, n-1$$ and 
suppose that the following conditions hold:
\roster
\item For any two subspaces $L_2 \subset L_1 \subset {\Bbb R}^n$ such that $\dim L_1 = m \leq n-1$ and $\dim L_2=m-1 \geq 0$, we have 
$$\left| D\left(p_1| L_1, \ldots, p_m|L_1\right) \right| \ \geq \ m \mu \left|D\left(p_1|L_2, \ldots, p_{m-1}|L_2 \right)\right|,$$
where we agree that for $m=1$ the inequality reads as 
$$\left|D\left(p_1|L_1\right)\right| \ \geq \ \mu;$$
\item We have 
$$|q_k(x)| \ \leq \delta \|x\|^2 \quad \text{for all} \quad x \in {\Bbb R}^n \quad \text{and all} \quad k=1, \ldots, n-1.$$
\endroster
Then for any two unit vectors $u, v \in {\Bbb R}^n$, we have 
$$D\left(p_1|u^{\bot}, \ldots, p_{n-1}|u^{\bot}\right)=(1+\alpha(u, v)) D\left(p_1|v^{\bot}, \ldots, p_{n-1}|v^{\bot}\right)$$
for some $\alpha(u, v) \in {\Bbb C}$ such that 
$$| \alpha(u, v)| \leq 4 \delta \mu^{-1} + 8 \delta^2 \mu^{-2}.$$
\endproclaim
\demo{Proof} As in Section 2.3, 
let us construct the quadratic forms $q_k^u, q_k^v: {\Bbb R}^{n-1} \longrightarrow {\Bbb R}$ for $k=1, \ldots, n-1$ and the corresponding forms 
$p_k^u, p_k^v: {\Bbb R}^{n-1} \longrightarrow {\Bbb C}$. Clearly, 
$$p_k^u(x)=\|x\|^2 + z_k q_k^u(x) \quad \text{and} \quad p_k^v(x)=\|x\|^2 + z_k q_k^v(x) \quad \text{for} \quad k=1, \ldots, n-1$$
and
$$\split &D\left(p_1|u^{\bot}, \ldots, p_{n-1}|u^{\bot}\right) = D\left(p_1^u, \ldots, p_{n-1}^u \right) \\  &\text{and} \quad 
D\left(p_1|v^{\bot}, \ldots, p_{n-1}|v^{\bot}\right)=D\left(p_1^v, \ldots, p_{n-1}^v\right). \endsplit$$
Let 
$$r_k(x)=q_k^u(x)-q_k^v(x) \quad \text{and hence} \quad p_k^u(x)-p_k^v(x)=z_k r_k(x) \quad \text{for} \quad k=1, \ldots, n-1.$$
From condition (2) of the lemma, we have
$$|r_k(x)| \ \leq \ 2\delta \|x\|^2 \quad \text{for} \quad k=1, \ldots, n-1. \tag3.2.1$$
From Corollary 2.5,
$$\aligned  D\left(p_1^u, \ldots, p_{n-1}^u \right) =&D\left(p_1^v, \ldots, p_{n-1}^v\right) + \sum_{i=1}^{n-1} z_i D\left(r_i, p_k^v: \ k \ne i \right) \\& \qquad + 
\sum_{1 \leq i < j \leq n-1} z_i z_j D\left(r_i, r_j, p_k^v:\ k \ne i, j \right) \endaligned \tag3.2.2$$
and
$$\rk r_k \ \leq \ 2 \quad \text{for} \quad k=1, \ldots, n-1.$$
If $n=2$ then the second sum is absent in the right hand side of (3.2.2).

We can write
$$r_k(x)=\lambda_{k1} \langle w_{k1}, x \rangle^2 + \lambda_{k2} \langle w_{k2}, x \rangle^2$$
where $\lambda_{k1}$ and $\lambda_{k2}$ are eigenvalues of $r_k$ with the corresponding unit eigenvectors $w_{k1}$ and $w_{k2}$. 
By (3.2.1) we have 
$$|\lambda_{k1}|,\ |\lambda_{k2}| \ \leq \  2\delta \quad \text{for} \quad k=1, \ldots, n-1.$$
Applying Lemma 2.2, we obtain
$$\split \left| D\left(r_i, p_k^v:\ k \ne i \right) \right| =& \left| \lambda_{i1} D\left(p_k^v| w_{i1}^{\bot}:\ k \ne i \right) + \lambda_{i2} D \left(p_k^v| w_{i2}^{\bot}:\ k \ne i \right) \right| \ 
\\ & \leq 2 \delta \left|  D\left(p_k^v| w_{i1}^{\bot}:\ k \ne i \right) \right| + 2 \delta \left|  D\left(p_k^v| w_{i2}^{\bot}:\ k \ne i \right) \right| \\
&\leq  4 \delta \max\Sb L \subset v^{\bot}: \\ \dim L = n-2 \endSb |D\left( p_k| L: \ k \ne i \right)| \\ &\leq {4 \delta \over (n-1) \mu} \left|D\left(p_1|v^{\bot}, \ldots, p_{n-1}|v^{\bot}\right)\right|, \endsplit$$
where the final inequality follows by condition (1) of the lemma.
For $n=2$ we just have 
$$\left| D\left(r_1\right)\right| \ \leq \ 2 \delta \ < \ {4 \delta \over \mu}.$$

Similarly, if $n \geq 3$, for every $1 \leq i < j \leq n-1$, we obtain 
$$\split \left| D\left(r_i, r_j , p_k^v:\ k \ne i, j \right) \right| 
\ \leq &\left| \lambda_{i1} \lambda_{j1}  D\left(p_k^v| w_{i1}^{\bot} \cap w_{j1}^{\bot} :\ k \ne i, j \right) \right| \\&+
\left| \lambda_{i1} \lambda_{j2} D\left(p_k^v| w_{i1}^{\bot} \cap w_{j2}^{\bot}: \ k \ \ne i, j \right)\right| 
\\ &+ \left| \lambda_{i2} \lambda_{j1}  D\left(p_k^v| w_{i2}^{\bot} \cap w_{j1}^{\bot} :\ k \ne i, j \right) \right| \\ &+
\left| \lambda_{i2} \lambda_{j2} D\left(p_k^v| w_{i2}^{\bot} \cap w_{j2}^{\bot}: \ k \ne i, j \right) \right|, \endsplit $$
where each of the four terms in the right hand side is $0$ if the corresponding intersection of subspaces $w_{i1}^{\bot}, w_{i2}^{\bot}, w_{j1}^{\bot}$ or $w_{j2}^{\bot}$ fails to be 
$(n-3)$-dimensional. Hence we get 
$$\split \left| D\left(r_i, r_j , p_k^v:\ k \ne i, j \right) \right| 
&\leq  16\delta^2 \max\Sb L \subset v^{\bot}: \\ \dim L = n-3 \endSb |D\left( p_k| L: \ k \ne i, j\right)| \\ &\leq {16 \delta^2  \over (n-1)(n-2) \mu^2} \left|D\left(p_1|v^{\bot}, \ldots, p_{n-1}|v^{\bot}\right)\right|, \endsplit$$
where for $n=3$ we just have 
$$\left| D\left(r_1, r_2\right) \right| \ \leq \ D\left(2 \delta I,\  2\delta I\right)=8\delta^2 \ < \ {16 \delta^2 \over 2 \mu^2}. $$

Hence
$$\left| \sum_{i=1}^{n-1} z_i D\left(r_i, p_k^v: k \ne i \right) \right| \ \leq \ 4 \delta \mu^{-1} \left| D \left(p_1|v^{\bot}, \ldots, p_{n-1}|v^{\bot}\right)\right|$$
and 
$$\left| \sum_{1 \leq i < j \leq n-1} z_i z_j  D\left(r_i, r_j, p_k^v:\ k \ne i, j \right)\right| \ \leq \ 8 \delta^2 \mu^{-2}\left| D \left(p_1|v^{\bot}, \ldots, p_{n-1}|v^{\bot}\right)\right|.$$ 
Therefore, from (3.2.2) we obtain
$$\split &\left| D\left(p_1|u^{\bot}, \ldots, p_{n-1}|u^{\bot}\right) -D\left(p_1|v^{\bot}, \ldots, p_{n-1}|v^{\bot}\right)\right| \\ &\quad  \leq \
(4 \delta \mu^{-1}+ 8 \delta^2 \mu^{-2}) \left| D \left(p_1|v^{\bot}, \ldots, p_{n-1}|v^{\bot}\right)\right| \endsplit$$
and the proof follows.
{\hfill \hfill \hfill}\qed
\enddemo

\subhead (3.3) Proof of Theorem 1.3 \endsubhead 
Let us choose $0 < \delta, \tau <1$ such that the following inequalities hold:
$$\delta + \tau + \delta \tau < 1 \quad \text{and} \quad {4 \delta \over 1- \delta -\tau-\delta \tau} + {8 \delta^2 \over (1-\delta - \tau -\delta \tau)^2} \ \leq \ \tau.$$
It is clear that for any $0 < \tau <1 $ one can find a sufficiently small $\delta >0$ such that the above inequalities are satisfied. We are interested to choose $\delta$ as 
large as possible and a computation shows that we can choose $\delta=0.045$ and $\tau=0.4$.

Let 
$$\mu=1-\delta -\tau - \delta \tau > 0.$$
As before, we introduce quadratic forms $q_k: {\Bbb R}^n \longrightarrow {\Bbb R}$ with matrices $Q_k$ and 
$p_k: {\Bbb R}^n \longrightarrow {\Bbb C}$ where
$$p_k(x) = \|x\|^2 + z_k q_k(x) \quad \text{for} \quad k=1, \ldots, n.$$
Suppose that 
$$|q_k(x)| \ \leq \ \delta \|x\|^2 \quad \text{for all} \quad x \in {\Bbb R}^n \quad \text{and} \quad k=1, \ldots, n.$$
We prove by induction on $n$ the following statements (3.3.1.$n$)--(3.3.3.$n$).
\bigskip
For any $p_1, \ldots, p_n: {\Bbb R}^n \longrightarrow {\Bbb C}$ as above, the following holds:
\medskip
(3.3.1.$n$) We have 
$$D\left(p_1, \ldots, p_n \right) \ne 0;$$
\medskip
(3.3.2.$n$) Suppose that $n \geq 2$. Then for any unit vectors $u, v \in {\Bbb R}^n$, we have 
$$D\left(p_1|u^{\bot}, \ldots, p_{n-1}|u^{\bot} \right) = \left(1+\alpha(u, v)\right) D\left(p_1|v^{\bot}, \ldots, p_{n-1}|v^{\bot} \right)$$
for some $\alpha(u, v) \in {\Bbb C}$ such that $|\alpha(u, v)| \ \leq \ \tau$;
\medskip
(3.3.3.$n$) Suppose that $n \geq 1$. Then for any unit vector $u \in {\Bbb R}^n$, we have 
$$\left|D\left(p_1, \ldots, p_n \right)\right| \ \geq \ \mu n \left| D\left(p_1|u^{\bot}, \ldots, p_{n-1}|u^{\bot}\right)\right|,$$
where for $n=1$ the inequality reads
$$\left|D\left(p_1\right)\right| \ \geq \ \mu.$$
\bigskip
We note that for $n=1$, we have $|D(p_1)| \geq 1-\delta$, so (3.3.1.1) and (3.3.3.1) hold. If $n \geq 2$, the statements (3.3.3.$m$) for $m=1, \ldots, n-1$ and Lemma 3.2 imply the statement (3.3.2.$n$). Then the statement (3.3.2.$n$) and Lemma 3.1 imply the statement (3.3.3.$n$). Finally, the statement (3.3.3.$n$) and (3.3.1.$n-1$) imply 
the statement (3.3.1.$n$). 

Hence we can choose $\gamma_0=\delta$ and the proof follows.
{\hfill \hfill \hfill} \qed

\head 4. Proof of Theorem 1.4 \endhead

The interpolation method is based on the following simple lemma, see for example, Lemma 2.2.1 of \cite{Ba16}.
\proclaim{(4.1) Lemma} Let $g: {\Bbb C} \longrightarrow {\Bbb C}$ be a univariate polynomial of degree $n$ and suppose that $g(z) \ne 0$ provided $|z| < \beta$, where $\beta >1$ is a real number.
Let us choose a branch of $f(z)=\ln g(z)$ for $|z| \leq 1$ and let 
$$T_m(z)=f(0) + \sum_{k=1}^m {f^{(k)}(0) \over k!}  z^k \quad \text{for} \quad |z| \leq 1$$
be the Taylor polynomial of degree $m$ computed at $z=0$. Then
$$\left| f(1)-T_m(1)\right| \ \leq \ {n \over \beta^m(\beta-1) (m+1)}.$$
\endproclaim 
\demo{Proof} Let $\alpha_1, \ldots, \alpha_n$ be the roots of $g$, counting multiplicity, so 
$$g(z)=g(0)\prod_{i=1}^n \left(1-{z \over \alpha_i} \right) \quad \text{and} \quad |\alpha_i| \ \geq \  \beta \quad \text{for} \quad i=1, \ldots, n.$$
Hence
$$f(z)=\ln g(z) = f(0) + \sum_{i=1}^n \ln \left(1- {z \over \alpha_i} \right).$$
Approximating the logarithms by their Taylor polynomials, we get 
$$\ln \left(1- {1 \over \alpha_i}\right)=-\sum_{k=1}^m {1 \over k \alpha_i^k} + \eta_i \quad \text{for} \quad i=1, \ldots, n,$$
where
$$\left| \eta_i \right|= \left| - \sum_{k=m+1}^{\infty} {1 \over k \alpha_i^k} \right| \ \leq \ \left| {1 \over m+1} \sum_{k=m+1}^{\infty} {1 \over \beta^k} \right| =
{1 \over \beta^m  (\beta-1)(m+1)}.$$
Since
$$T_m(1)= -\sum_{i=1}^n \sum_{k=1}^m {1 \over k \alpha_i^k},$$
the proof follows.
{\hfill \hfill \hfill} \qed
\enddemo
In particular, to approximate $f(1)$ within an additive error $0 < \epsilon < 1$, it suffices to choose $m=O(\ln n- \ln \epsilon)$, where the implicit constant in the ``$O$" notation depends on $\beta$ alone.

\subhead (4.2) Computing $f^{(k)}(0)$ \endsubhead Under the conditions of Lemma 4.1, to approximate $f(1)$ within an additive error $0 < \epsilon < 1$, it suffices to 
compute $f(0)$ and the derivatives $f^{(k)}(0)$ for $k=1, \ldots, m$ for some $m =O(\ln n - \ln \epsilon)$. That, in turn, can be reduced to computing $g(0)$ and 
$g^{(k)}(0)$ for $k=1, \ldots, m$. Indeed, we have 
$$f'(z)={g'(z) \over g(z)} \quad \text{from which} \quad g'(z) = f'(z) g(z).$$
Differentiating the product $k-1$ times, we obtain
$$g^{(k)}(0)=\sum_{j=0}^{k-1} {k-1 \choose j} f^{(k-j)}(0) g^{(j)}(0) \quad \text{for} \quad k=1, \ldots, m. \tag 4.2.1$$
We consider (4.2.1) as a triangular system of linear equations in the variables $f^{(k)}(0)$ for $k=1, \ldots, m$. The diagonal coefficients are $g^{(0)}(0)=g(0) \ne 0$, so the matrix of the system is invertible. Given the values of $g(0)$ and $g^{(k)}(0)$ for $k=1, \ldots, m$, one can compute the values of $f^{(k)}(0)$ for $k=1, \ldots, m$ in 
$O(m^2)$ time, see Section 2.2.2 of \cite{Ba16}.

Now we are ready to prove Theorem 1.4. 

\subhead (4.3) Proof of Theorem 1.4 \endsubhead Given matrices $Q_1, \ldots, Q_n$ and complex numbers $z_1, \ldots, z_n$, let us denote
$$A_k=z_k Q_k \quad \text{for} \quad k=1, \ldots, n$$ 
and let us define a univariate polynomial 
$$g(z)=D\left(I + zA_1, \ldots, I+zA_n \right).$$
Hence $\deg g(z) \leq n$ and by Theorem 1.3 we have 
$$g(z) \ne 0 \quad \text{provided} \quad |z| \ \leq \ {1 \over \rho} = \beta.$$
We define $f(z)=\ln g(z)$ for $|z| \leq 1$ and let $T_m(z)$ be the Taylor polynomial of $f(z)$ of degree $m$ computed at $z=0$. From Lemma 4.1, we conclude that 
$T_m(1)$ approximates $f(1)$ within an additive error of $0 < \epsilon < 1$ for some $m \leq c(\rho)(\ln n -\ln \epsilon)$. 

It remains to show that $f^{(k)}(0)$ is a polynomial 
in the entries of the matrices $A_1, \ldots, A_n$ of degree at most $k$. In view of  the equations (4.2.1), it suffices to show that $g^{(k)}(0)$ is a polynomial of degree at most 
$k$ in the entries of $A_1, \ldots, A_n$. Now, by the linearity of the mixed discriminant in each argument, we can write 
$$D\left(I + z A_1, \ldots, I + z A_n \right)=\sum_{k=0}^n  z^k \sum\Sb J \subset \{1, \ldots, n\} \\ |J|=k \endSb D\left(\underbrace{I, \ldots, I}_{\text{$n-k$ times}}, A_j: \ j \in J \right).$$
Hence 
$$\split g(0)=&D(I, \ldots, I)=n! \quad \text{and} \\  g^{(k)}(0)=&k! \sum\Sb J \subset \{1, \ldots, n\} \\ |J|=k \endSb D\left(\underbrace{I, \ldots, I}_{\text{$n-k$ times}}, A_j: \ j \in J \right). \endsplit  \tag4.3.1$$
Denoting $A_s=\left(a_{ij}^s\right)$, we obtain from (1.1.1) that
$$D\left(\underbrace{I, \ldots, I}_{\text{$n-k$ times}}, A_j: \ j \in J \right)= \sum\Sb \sigma, \tau \in S_n \\ \sigma(j)=\tau(j) \text{ \ for \ } j \notin J \endSb \sgn(\sigma \tau) \prod_{j \in J} a^j_{\sigma(j) \tau(j)}. \tag4.3.2$$
It follows from (4.3.2) that $g^{(k)}(0)$ is indeed a homogeneous polynomial of degree $k$ in the entries of $A_1, \ldots, A_n$, which concludes the proof.
{\hfill \hfill \hfill} \qed

We can rewrite the right hand side of (4.3.2) as 
$$(n-k)! \sum\Sb \phi, \psi: J \longrightarrow \{1, \ldots, n\} \\ \phi(J)=\psi(J) \endSb \sgn(\phi \psi) \prod_{j \in J} a^j_{\phi(j) \psi(j)},$$
where the sum is taken over all ${n \choose k} (k!)^2$ pairs of injections $\phi, \psi: J \longrightarrow \{1, \ldots, n\}$ such that $\phi(J)=\psi(J)$ and $\sgn(\psi \phi)=\sgn(\sigma\tau)$, where $\sigma, \tau \in S_n$ are any two permutations such that the permutation $\sigma$ agrees with $\phi$ on $J$, the permutation $\tau$ agrees with $\psi$ on $J$ and $\sigma$ agrees with $\tau$ outside of $J$ (as is easy to see, $\sgn(\sigma \tau)$ does not depend on a particular choice of $\sigma$ and $\tau$).
It follows then that $g^{(k)}(0)$ and hence $f^{(k)}(0)$ can be computed for $k=1, \ldots, m$ in $n^{O(m)}$ time. Since we choose $m=O(\ln n - \ln \epsilon)$ we obtain a quasi-polynomial algorithm for computing the approximation $p(Q_1, \ldots, Q_n; z_1, \ldots, z_n)$.

\head 5. Proofs of Theorems 1.5 and 1.6 \endhead

\subhead (5.1) Mixed characteristic polynomial and mixed discriminants \endsubhead Let $p(x)$ be the mixed characteristic polynomial of $n \times n$ matrices 
$A_1, \ldots, A_m$ defined by (1.1.4). Our immediate goal is to express the coefficients of $p(x)$ in terms of mixed discriminants.
Let 
$$Q=Q\left(x, z_1, \ldots, z_n\right)=xI + \sum_{k=1}^m z_k A_k.$$
Using the linearity of the mixed discriminant in each argument, we obtain
$$\split \det Q={1 \over n!} D\left(Q, \ldots, Q\right) =& {1 \over n!} \sum_{k=0}^n x^k {n \choose k}   \sum\Sb r_1, \ldots, r_m \geq 0 \\ r_1 + \ldots + r_m =n-k \endSb {n-k \choose  r_1, \ldots,  r_m} z_1^{r_1} \cdots z_m^{r_m}  \\ &\quad \times D\left(\underbrace{I, \ldots, I}_{\text{$k$ times}}, \underbrace{A_1, \ldots, A_1}_{\text{ $r_1$ times}}, \ldots, \underbrace{A_m, \ldots, A_m}_{\text{$r_m$ times}}\right). \endsplit$$
Therefore,
$$p(x)=(-1)^n \sum_{k=0}^n (-1)^k {x^k \over k!}  \sum\Sb J \subset \{1, \ldots, m\} \\ |J|=n-k \endSb D\left(\underbrace{I,\ldots, I}_{\text{$k$ times}}, A_j: j \in J \right). \tag5.1.1$$

Next, we need two results of Szeg\H{o}. The first result concerns operations on polynomials with no zeros in a disc.
\proclaim{(5.2) Theorem} Let $q, r: {\Bbb C} \longrightarrow {\Bbb C}$ be univariate polynomials, 
$$q(z)=\sum_{k=0}^n a_k z^k \quad \text{and} \quad r(z)=\sum_{k=0}^n b_k z^k$$
and let us define $s=q \star r$ by
$$s(z)=\sum_{k=0}^n {a_k b_k \over {n \choose k}} z^k.$$
Suppose that $q(z) \ne 0$ whenever $|z| \leq \lambda$ and $r(z) \ne 0$ whenever $|z| \leq \mu$ for some $\lambda, \mu >0$.
Then $s(z) \ne 0$ provided $|z| \le \lambda \mu$.
\endproclaim
For the proof, see, for example, Corollary 2.5.10 in \cite{Ba16}. 

The second result concerns the complex roots of a particular polynomial, see for example, Lemma 5.5.4 of \cite{Ba16}.
\proclaim{(5.3) Theorem} Let 
$$r(z) = \sum_{k=0}^n {z^k \over k!}.$$
Then $r(z) \ne 0$ whenever $|z| \leq \alpha_0 n$, where $\alpha_0 \approx 0.278$ is the positive real solution of the equation $\alpha e^{1+\alpha} =1$.
\endproclaim

In their proof of the Kadison-Singer Conjecture, Marcus, Spielman and Srivastava obtained the following crucial result \cite{M+15}, see also \cite{MS17}.
\proclaim{(5.4) Theorem} Let $A_1, \ldots, A_m$ be positive semidefinite Hermitian matrices such that 
$$\sum_{k=1}^m A_k = I \quad \text{and} \quad \tr\left(A_k\right) \ \leq \ \epsilon \quad \text{for} \quad k=1, \ldots, m$$
and some $\epsilon \geq 0$. Then the roots of the mixed characteristic polynomial 
$$p(x)=p_{A_1, \ldots, A_m}(x)=  \prod_{k=1}^m \left(1 -{\partial \over \partial z_k}\right) \det\left(xI + \sum_{k=1}^m z_k A_k \right)\Big|_{z_1= \ldots = z_m=0}$$
are non-negative real and do not exceed $(1+\sqrt{\epsilon})^2$.
\endproclaim
More precisely, real-rootedness of the mixed characteristic polynomial is Corollary 4.4 of \cite{M+15}, the upper bound on the roots is Theorem 5.1 of \cite{M+15}, while non-negativity of the roots follows,
for example, from the fact that the coefficients of the polynomial (5.1.1) alternate in sign.

Now we are ready to prove Theorem 1.5.
\subhead (5.5) Proof of Theorem 1.5 \endsubhead We consider the mixed characteristic polynomial $p(z)$ of the matrices $Q_1, \ldots, Q_n$. By (5.1.1), we have 
$$p(z)=(-1)^n \sum_{k=0}^n (-1)^k {z^k \over  k!} \sum\Sb J \subset \{1, \ldots, n\} \\ |J|=n-k \endSb D\left(\underbrace{I, \ldots. I}_{\text{$k$ times}}, Q_j:  j \in J \right).$$
Choosing $m=n$ and $\epsilon=1$ in Theorem 5.4, we conclude that $p(z) \ne 0$ provided $|z| > 4$.

Let us define
$$q(z)=z^n p\left({1 \over z}\right)=\sum_{k=0}^n (-1)^k {z^k \over (n-k)!} \sum\Sb J \subset \{1, \ldots, n\} \\ |J|=k \endSb D\left(\underbrace{I,\ldots, I}_{\text{$n-k$ times}}, Q_j: j \in J \right).$$
Hence $q(z)$ is a polynomial of degree $n$ and $q(z) \ne 0$ provided $|z| < 1/4$.

Let 
$$r(z)=\sum_{k=0}^n {z^k \over k!}$$ 
and let 
$$\split s(z)=&q(z) \star r(z)={1 \over n!} \sum_{k=0}^n (-1)^k z^k \sum\Sb J \subset \{1, \ldots, n\} \\ |J|=k \endSb
 D\left(\underbrace{I,\ldots, I}_{\text{$n-k$ times}}, Q_j: j \in J \right) \\=
&{1 \over n!} D\left(I-zQ_1, \ldots, I- zQ_n \right). \endsplit$$
Applying Theorems 5.2 and 5.3, we conclude that $s(z) \ne 0$ provided $|z| < \alpha_0 n/4$, which completes the proof.
{\hfill \hfill \hfill} \qed

The proof of Theorem 1.6 is completely similar to that of Theorem 1.4 in Section 4.3 and therefore omitted.

\head 6. Proofs of Theorems 1.7 and 1.8 \endhead

\definition{(6.1) Definition} Let 
$${\Bbb D}=\left\{ z \in {\Bbb C}:\ |z| \ \leq 1 \right\}$$ be the unit disc in the complex plane. We say that an $n$-variate polynomial $p\left(z_1, \ldots, z_n\right)$ is {\it ${\Bbb D}$-stable}
if 
$$p\left(z_1, \ldots, z_n\right) \ne 0 \quad \text{provided} \quad z_1, \ldots, z_n \in {\Bbb D}.$$
\enddefinition
We are interested in {\it multi-affine} $n$-variate polynomials, where the degree in each variable does not exceed $1$. For $S \subset \{1, \ldots, n\}$, we denote
$$\zz^S =\prod_{i\in S} z_i,$$
where we agree that $\zz^{\emptyset}=1$. Hence a multi-affine $n$-variate polynomial can be written as 
$$p\left(z_1, \ldots, z_n\right)=\sum_{S \subset \{1, \ldots, n\}} a_S \zz^S,$$
where $a_S$ are complex coefficients.

The following result, known as {\it Asano contractions}, asserts that the Schur-Hadamard product of ${\Bbb D}$-stable multi-affine polynomials is ${\Bbb D}$-stable, see, for example, Theorem 2.5.1 of \cite{Ba16}.
\proclaim{(6.2) Theorem} Suppose that $n$-variate polynomials 
$$f\left(z_1, \ldots, z_n\right)=\sum_{S \subset \{1, \ldots, n\}} a_S \zz^S \quad \text{and} \quad g\left(z_1, \ldots, z_n \right)=\sum_{S \subset \{1, \ldots, n\}} b_S \zz^S $$
are ${\Bbb D}$-stable. Then the polynomial $h=f \ast g$ defined by 
$$h\left(z_1, \ldots, z_n\right)=\sum_{S \subset \{1, \ldots, n\}} a_S b_S \zz^S $$
is also ${\Bbb D}$-stable.
\endproclaim

\subhead (6.3) Proof of Theorem 1.7 \endsubhead For $\zz=\left(z_1, \ldots, z_n\right)$, let $D(\zz)$ be the $n \times n$ diagonal matrix matrix having $z_1, \ldots, z_n$ on the diagonal. 
Then for any $z_1, \ldots, z_n \in {\Bbb D}$, we have 
$$\| D(\zz) B x \| \ \leq \ \|B x\| \ < \ \|x\| \quad \text{for every vector} \quad x \in {\Bbb C}^n \setminus \{0\}.$$
Therefore, 
$$\ker\left(I + D(\zz) B \right)=\{0\},$$ the matrix $I + D(\zz) B$ is invertible and the polynomial 
$$\det \left( I + D(\zz) B \right) =\sum_{S \subset \{1, \ldots, n\}} \left( \det B_S\right) \zz^S$$
is ${\Bbb D}$-stable. Applying $m$ times Theorem 6.2, we conclude that the polynomial 
$$h_m\left(z_1, \ldots, z_n \right) = \sum_{S \subset \{1, \ldots, n\}} \left(\det B_S \right)^m \zz^S$$
is ${\Bbb D}$-stable and hence 
$$\phi_{B, m}(z) = h_m\left(z, \ldots, z\right) \ne 0 \quad \text{provided} \quad |z| \leq 1,$$
as required.
{\hfill \hfill \hfill} \qed

\subhead (6.4) Proof of Theorem 1.8 \endsubhead 
Since $\|B\| < \rho$, we have $\| \rho^{-1} B \| < 1$ and hence by Theorem 1.7,
$$\split \phi_{B,m}(z)= &\sum_{S \subset \{1, \ldots, n\}} \left(\det B_S\right)^m z^{|S|}=\sum_{S \subset \{1, \ldots, n\}} \left( \det \left(\rho^{-1} B\right)_S\right)^m \left(z \rho^m\right)^{|S|}\\=
&\phi_{\rho^{-1}B, m} \left(\rho^m z \right) \ne 0 \quad \text{provided} \quad |z| \leq \rho^{-m}. \endsplit$$
We define $f(z)=\ln \phi_{B,m}(z)$ for $|z| \leq 1$ and let $T_m(z)$ be the Taylor polynomial of $f(z)$ of degree $m$ computed at $z=0$. From Lemma 4.1, we conclude that 
$T_m(1)$ approximates $f(1)$ within an additive error $0 < \epsilon < 1$ for some $m \leq c(\rho)\left(\ln n - \ln \epsilon\right)$. We observe that $\phi_{B,m}^{(k)}(0)$ is a polynomial in the entries of $B$ of degree $km$ which can be computed in $n^{O(km)}$ time. From Section 4.2, it now follows that $f^{(k)}(0)$ is a polynomial in the entries of $B$ of degree $km$ which can be computed in $n^{O(km)}$ time.
{\hfill \hfill \hfill} \qed

\head Acknowledgments \endhead

I am grateful to Shayan Oveis Gharan for pointing out to the problem of computing (1.1.5) and to \cite{KT12}, and to him, Leonid Gurvits, Jingcheng Liu, Mohan Ravichandran and Piyush Srivastava for many inspiring conversations about mixed discriminants during the ``Geometry of Polynomials" program at the Simons Institute for the Theory of Computing.

\enddocument
\end